\documentstyle[11pt,a4,psfig,fullpage]{article}
\newcommand{\be}{\begin{equation}}
\newcommand{\ee}{\end{equation}}
\newcommand{\beq}{\begin{eqnarray}}
\newcommand{\eeq}{\end{eqnarray}}

\begin{document}
\pagestyle{empty}
\begin{flushright}
{BROWN-HET-1184} \\
{IHEP 99-32} \\
{KIAS-P99070} \\
{ULG-PNT-99-JRC-1} \\
{Revised November 2000} \\
\end{flushright}
\vspace*{5mm}

\begin{center}
{\Large High-Energy Forward Scattering and the Pomeron:\\
Simple Pole versus Unitarized Models}\\ [10mm]

J.R. Cudell$^a$\footnote{JR.Cudell@ulg.ac.be}, V.
Ezhela$^b$\footnote{ezhela@mx.ihep.su},
 K. Kang$^c$\footnote{kang@het.brown.edu;
supported in part by DOE Grant DE-FG02-91ER40688 - Task
A}, S. Lugovsky$^b$\footnote{lugovsky@mx.ihep.su}, and N.
Tkachenko$^{b}$\footnote{tkachenkon@mx.ihep.su}\\
[5mm]
{\em a.  Inst. de Physique, Universit\'e de Li\`ege, B$\hat{a}$t. B-5,
Sart Tilman, \\B4000 Li\`ege, Belgium} \\
[5mm]
{\em b.  COMPAS Group\footnote{
 This work of the COMPAS Group is supported in part by RFBR Grants
RFBR-96-07-89230 and RFBR-99-07-90356. }, IHEP, Protvino, Russia}\\
[5mm]
{\em c.  Department of Physics, Brown University, Providence RI 02912 USA
and  School of
Physics, Korea Institute for Advanced Study, Seoul 130-012, Korea}\\
[15mm]

Abstract
\end{center}
\begin{quote}
Using the largest data set available, we determine the best values that
the data at $t=0$ (total cross sections
and real parts of the hadronic amplitudes) give for the intercepts
and couplings of the soft pomeron and of the $\rho/\omega$ and $a/f$
trajectories.
We show that these data cannot discriminate between a simple-pole fit
and asymptotic $\log^2s$ and $\log s$ fits, and hence are not sufficient
to reveal the ultimate nature of the pomeron. However, we evaluate the existing
evidence (factorization, universality, quark counting) favouring
the simple-pole hypothesis.
We also examine the range of validity in energy of the fits, and
show that one cannot rely on such fits in the region $\sqrt{s}<9$ GeV. We
also establish
bounds on the odderon and the hard pomeron.
\end{quote}
\vspace*{3mm}

\newpage
\setcounter{page}{1}
\pagestyle{plain}

\section*{Introduction}
The description of forward scattering by universal fits has been an open
question for the last twenty years. The data from HERA, which now extend
the measurement of off-shell cross sections to very low values of $Q^2$,
have revived the interest in this problem, as it
can shed some light on the nature of the pomeron.
Because of the presence of large logarithms of the center-of-mass energy
$\sqrt{s}$, perturbative QCD predicts an explosive increase of the cross
sections with energy. Whether this prediction is stable remains to be
seen, but such a
sharp rise is qualitatively present in the DIS data
from HERA. However, this is in
marked contrast with the observation of
on-shell hadronic total cross sections, which have a very slow rise
with $s$.

Two schools of
thought exist regarding this puzzle. The first one starts from
the simplest assumption
within
Regge theory: that this rise with $s$ is the result of
the presence of a glueball tra\-jectory, for which there are
at present strong candidates
\cite{glueballs}. This trajectory is called
the pomeron, and has an intercept
slightly larger than 1. This assumption leads to the prediction of
a universal rise with $s$, and of factorization.
The further hypothesis that the
pomeron couples to constituent quarks leads further to the prediction
of quark-counting rules. Moreover, simple refinements have enabled Donnachie
and Landshoff (DL)
to push these ideas further  \cite{DoLa}, and
to reproduce qualitatively well all soft data
for the scattering of on-shell particles, even at non-zero $t$.
The problem with this approach is that it cannot be automatically extended
to off-shell particles, and, in particular, to DIS.
The only possible hypothesis \cite{twopoms}
would be that an
extra trajectory enters the problem, and that this trajectory
decouples at $Q^2=0$.
The possibility of such a stable trajectory is phenomenologically
viable, and is confirmed,
to some extent, by the DGLAP evolution.

The other school of thought starts from
perturbative QCD, and assumes that
unitarization changes the fierce rise observed at large $Q^2$ to
something compatible
with the Froissart bound. This approach suffers from the fact that, despite
recent progress \cite{Lam}, no one has reliably unitarized a QCD cross section.
However, it is clear that such a unitarization will involve the exchange
of a very large number of gluons between the quarks. Hence, the details
of the quark structure -- the hadronic wave function -- should matter, and this
means that the simple Regge factorization property would be lost, as well
as quark counting and even strict universality \cite{CN}.
Furthermore, it is expected that such a unitarization would
lead to a cut singularity instead of a pole, and
to a power behaviour in $\log s$.

The question we want to address here is whether one can distinguish
between these
two approaches by studying soft data. In order to maximize the number of
data points, we shall consider the full hadronic amplitude, {\it i.e.}
both the total cross section, giving the imaginary part,
and the $\rho$
parameter, giving the ratio of the real part to the imaginary part.
As we shall see,
the consideration of the $\chi^2$ alone surprisingly does
not discriminate between the different hypotheses, but leads
one to refine the description of lower trajectories, and to
define a minimum energy below which none of these fits work. The only
discrimination that the soft data can bring in lies in the confirmation
of the properties that suggest that the pomeron is a simple pole
coupled to
the
constituent quarks, {\it i.e.} universality, factorization and quark
counting.

This study complements and expands
the results of a recent letter
\cite{six}, where two of us (JRC and KK) with
S. K. Kim presented
a detailed statistical analysis of the parameters of the DL model
(\cite{six}),
as well as the analysis subsequently presented (by VE, SL and NT) in the 1998
Review of particle physics \cite{PDG}.

This paper is organized as follows. In section 1, we describe the
data sample and the hypothesis-testing procedure. In section 2,
we concentrate on the
simple-pole fit, and study first the changes one has to
introduce in order
to describe the low-energy data reliably. In section 3, we present the
evidence for simple-pole behaviour, and in section 4 we consider alternative
(unitary) forms for the pomeron-exchange term. In section 5, we mention
several attempts to extend the fit to the low-energy region. In section 6, we
use our dataset
to place bounds on other trajectories,
and we present some
predictions
for cross sections.

\section{Dataset and statistical procedure}
\subsection{Dataset}
Three of us (VE, SL and NT)
have prepared a complete and maintained \cite{dataset}
 set of published data for the cross sections and
real-to-imaginary part ratios for the following processes: $pp$, $\bar p p$,
$\pi^\pm p$
and $K^\pm p$, as well as for the total cross sections of $\gamma p$ and
$\gamma\gamma$ scattering. Some superseded points have been removed,
 and typos have been corrected.
It was found in \cite{six} that irrespectively of the models used, the $\,
\chi^2$/d.o.f. was large due to bad data points at ISR energies. Once
about 10\% of the ISR points were removed, an acceptable $\, \chi^2$/d.o.f.
was achieved,
leading to reliable estimates
 of the
parameters and their errors.  As it will turn out, the new dataset does not
necessitate such a filtering procedure, and thus seems more coherent.
The dataset contains 2747 (303) data points for total cross
sections (resp. real-to-imaginary ratios), and
the number of points
used in the fits is
given as a function of energy in
Fig.~1.
It is our hope that this dataset
will become the standard reference when studying the validity of models
for forward quantities.
\subsection{Definition of $\chi^2$ and of the errors}
As the dataset is quite large and has no substantial inconsistencies,
the conventional definition of $\chi^2$ is used. Note, however, that as the
most interesting
quantities are sensitive to the highest-$s$ region, where the data are scarce,
our definition may not be the best suited
for determining
the soft pomeron intercept and other definitions giving more weight
to the highest-energy data are possible.
One may also worry about whether one should consider only total cross
sections, or the full amplitude.  The best data are certainly the
measurements of the total cross section, and one might wonder  whether the
interference between pomeron exchange and the Coulomb cross sections can be
reliably calculated.

Despite these two worries, we see no fundamental reason for rejecting
part of
the data, or using a non-conventional $\chi^2$, as was done in
{\it e.g.} \cite{twopoms,DL}, where equal
weights were given to the $\bar p p$  and the $pp$ datasets,
while not fitting to the other cross sections
or to the $\rho$ parameters.

Choosing a conventional $\chi^2$ definition and
weighting all the points
with inverse squares of their total errors
enables us to define errors through the usual definition\footnote{Note that the
errors in this paper are smaller than those
of \cite{six} because there a change of
5 units was considered, and because we
have now a larger dataset.} of a change of
$\chi^2$ of 1
unit for acceptable fits with a $\chi^2$/d.o.f. of order 1. In
case of bad fits, we shall sometimes give an estimate of the error, which
corresponds to a change of $\chi^2$ of $\chi^2_{min}$/d.o.f., in other word, we
shall then dilate the errors by
the Birge factor.
This definition also allows us to reject models
or parameters corresponding to values of the $\chi^2$/d.o.f. appreciably larger
than 1.
Note that for the total error, we have added the statistical and the systematic
errors in quadrature.

One further problem is linked to the fact that the fits considered
below
are asymptotic: it is clear that smooth functions cannot
describe the resonance
region,
hence the fits can be trusted only above a certain energy
$\sqrt{s_{min}}$ which is
a parameter in itself
and could, in principle, be process-dependent.
We demand for the fits to be trusted that the value of
the parameters remains stable w.r.t.  $\sqrt{s_{min}}$, and that the
$\chi^2$/d.o.f.
be
less than 1.
This criterion implies that our determination of the
parameters describing the pomeron is stable, or equivalently that
the low energy data are not of primary importance.

\section{Regge fits and lower trajectories}
\begin{figure}
\psfig{figure=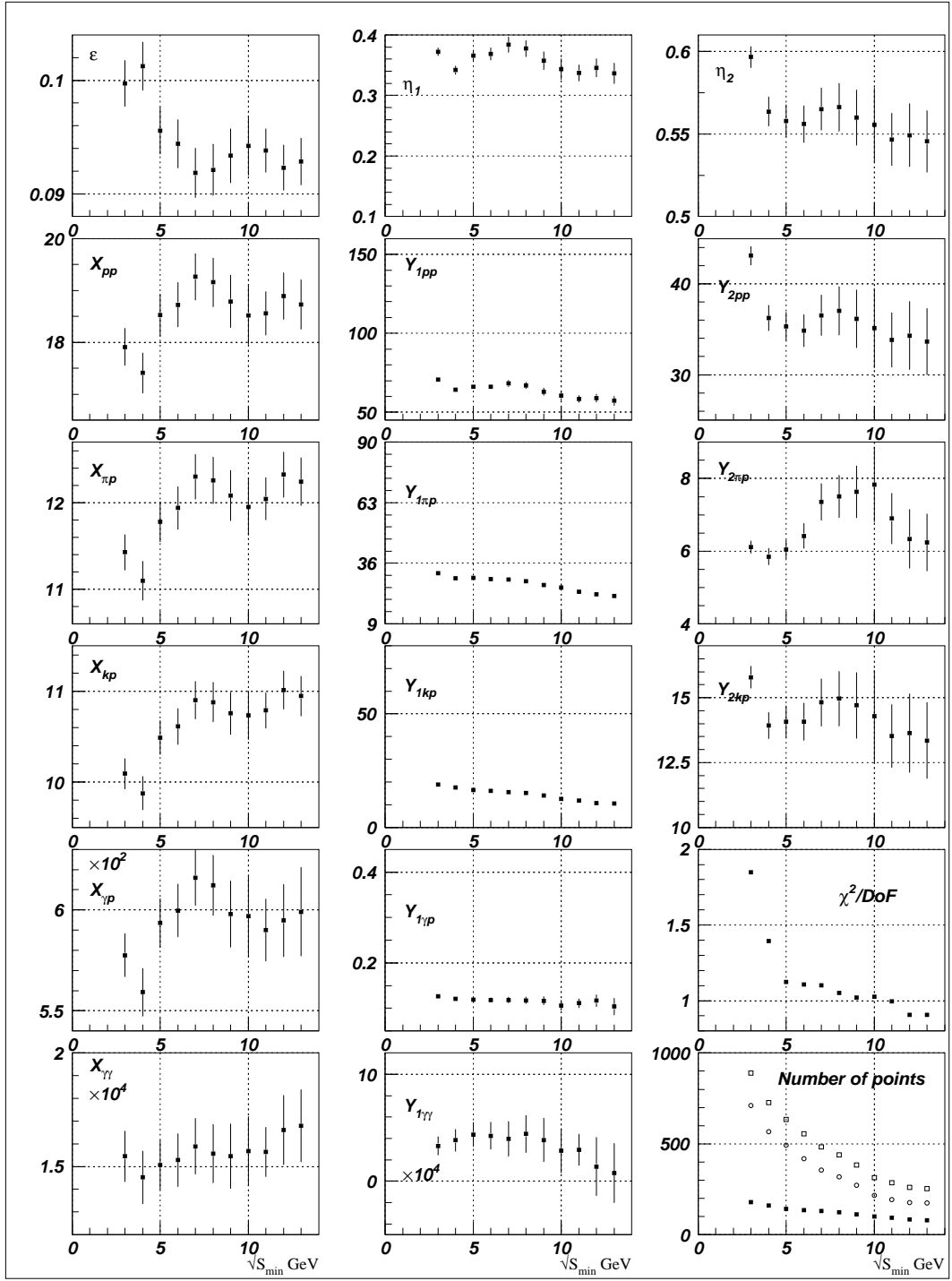,bbllx=0.7cm,bblly=1.3cm,bburx=20.3cm,bbury=28.3cm,width=14cm}
\caption{Parameters of the RRP model as functions of the minimum energy
considered in the fit.}\label{fig:RRP}
\end{figure}
\begin{figure}
\psfig{figure=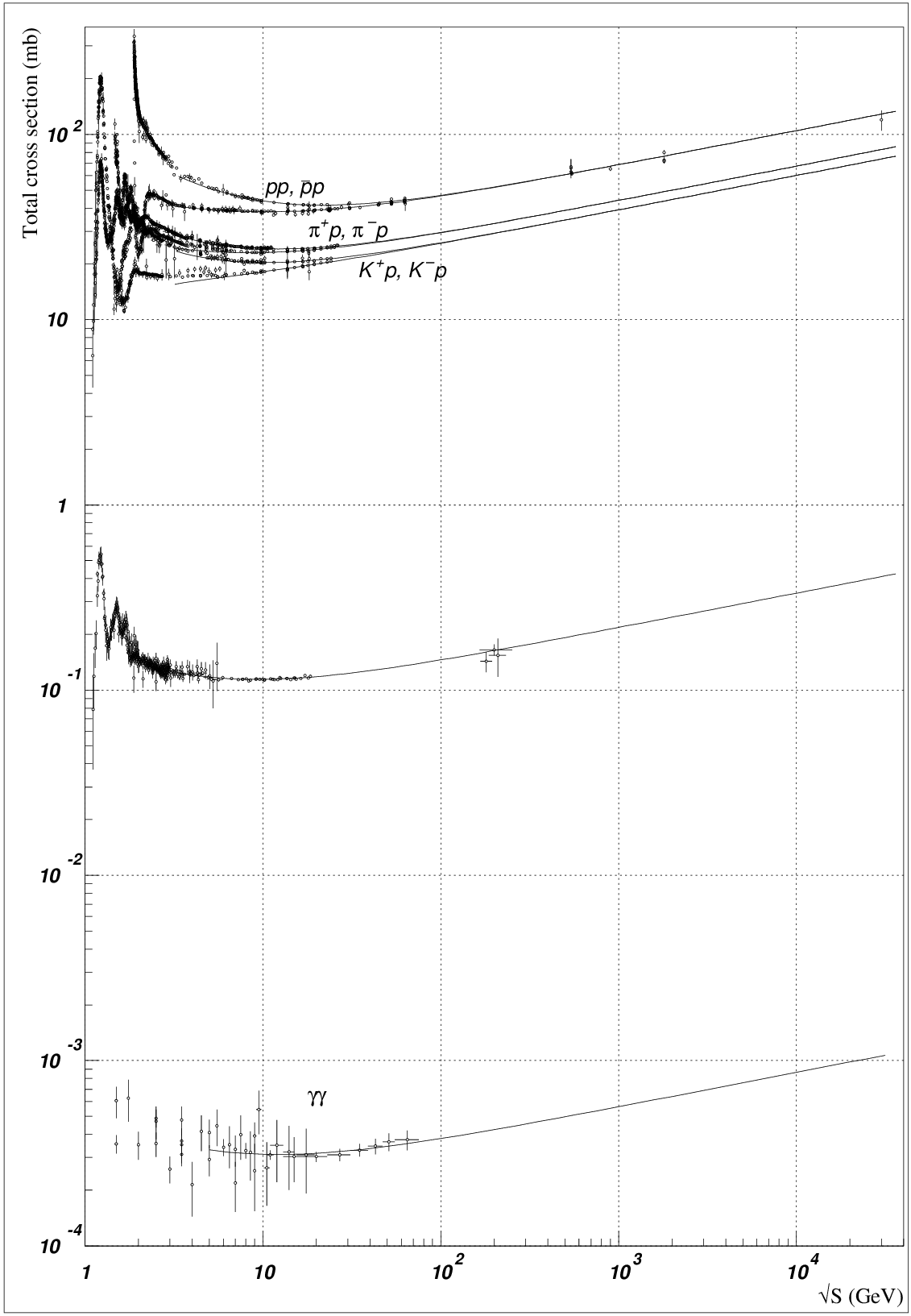,bbllx=0.7cm,bblly=1.3cm,bburx=20.3cm,bbury=28.3cm,width=14cm}
\vglue -10.5cm
\centerline{$\gamma p$}
\vglue +10.0 cm
\caption{The fit to the total cross sections from the
parametrization RRP.
}
\label{fig:RRPsig}
\end{figure}
\begin{figure}
\psfig{figure=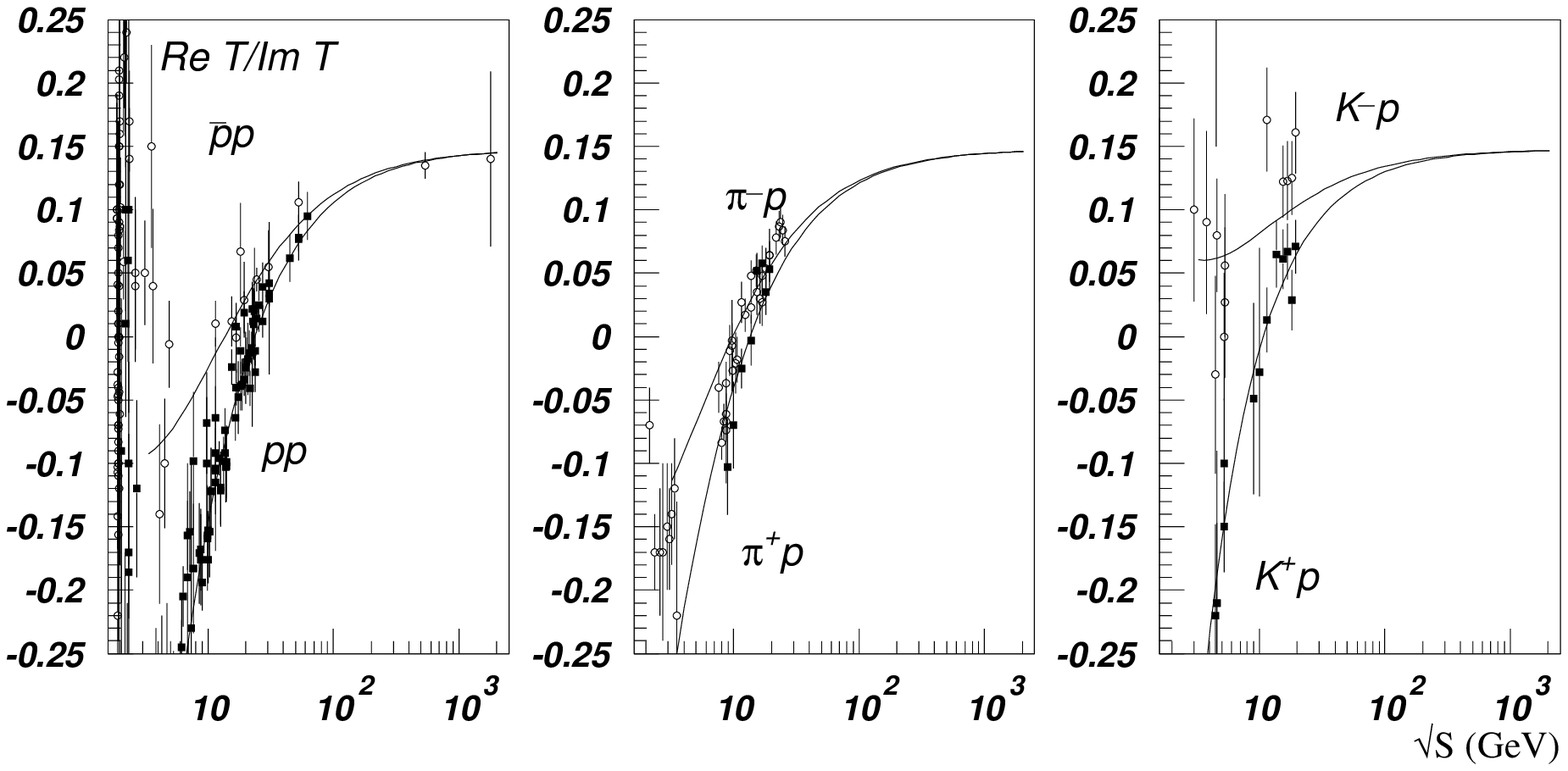,bbllx=1.3cm,bblly=10.9cm,bburx=20.cm,bbury=19.7cm,width=14cm}
\caption{The fit to the $\rho$ values from
the parametrization RRP.
}
\label{fig:RRPrho}
\end{figure}

First we discuss the
Regge-pole parametrization of the data.
It is based on the idea that the cross sections should be reproduced by
the simplest singularities in the complex $J$ plane, {\it i.e.} simple poles,
corresponding to the exchange of bound-state trajectories.
The imaginary part of the hadronic amplitude is then given by
\be
{\it ImA}_{h_1h_2}(s,t)=\sum_i (\pm 1)^{S_i} C_{h_1h_2}(t)
\left({s\over s_0}\right)^{\alpha_i(t)}
\label{imamp}
\ee
with $S_i$, the signature of the exchange. The total cross section
is
then equal to
\be
\sigma_{tot} (s) = Im\, A (s,0)/s .
\label{stot}
\ee
The trajectories $\alpha_i(t)$ are universal, and the process (and mass)
dependence is present only in the constants $C_{h_1h_2}(0)$
(which absorb the scale $s_0$). The highest trajectory, responsible for the
rise
of cross sections, is that of the soft pomeron. The others are those of the
mesons, and, in principle,
they are numerous. However, once the energy is high
enough, only those with the largest intercept, $\alpha_i(0)$
of order $1/2$, will contribute at $t=0$.
The four highest meson
trajectories can be clearly seen in a $M^2$ vs. $J$ plot of the meson data.
They correspond to the $\rho$, the $a$, the $\omega$, and the $f$
resonant states.

The simplest assumption, which would result from a simple string model
of the mesons, is that these trajectories are degenerate, which
implies that they have
the same intercepts \cite{DL}.
However, the results of a previous fit \cite{six} show that
an exchange-degenerate meson
trajectory fails to satisfy the proposed criteria: the $\chi^2$/d.o.f.
is large (of order 1.3), the parameters and their
errors are unstable when the model is fitted to the total cross sections and
the $\rho$ parameter, and, in
fact, the assumption of exchange degeneracy for $\, C=\pm 1\,$ meson
trajectories is not supported even by fits to total cross sections
only.

This situation persists for other parametrizations of the pomeron term, and
in the following we shall keep the low-energy model of cross sections
presented here and resulting from the exchange of two non-degenerate $C=+1$ and
$C=-1$ meson trajectories.

Furthermore, the situation remains identical when one considers the additional
data
presented here. Hence, we shall adopt the simple generalization
proposed in \cite{six}, which we shall call the RRP  model, and
which assumes independent $C=+1$ ($a/f$) and $C=-1$ ($\rho/\omega$)
intercepts. Hence, the formula (\ref{imamp}) for the total cross section
becomes
\beq
{{\it Im A}_{h_1h_2}(s)\over s} &=& X^{h_1h_2} s^{\epsilon} +
Y^{h_1h_2}_1
s^{-\eta_{1}} \mp Y^{h_1h_2}_2 s^{-\eta_{2}}\eeq
with the intercepts given by
\beq
\alpha_P&=&1+\epsilon\nonumber\\
\alpha_{\small (C=+1)}&=&1-\eta_1\\
\alpha_{\small (C=-1)}&=&1-\eta_2\nonumber
\eeq
The sign of the $Y_2$ term flips
when fitting $h_1\bar h_2$
data compared to $h_1h_2$ data.
The real parts of the forward elastic
amplitudes are calculated from analyticity
(see, for example \cite{KK,Kang,block}):
\beq
{{\it Re A}_{h_1h_2}(s)\over s} &=& -X^{h_1h_2}  s^{\epsilon}
\mbox{cot} \left( \frac{1+\epsilon}{2}\pi\right)
- Y^{h_1h_2}_1  s^{-\eta_{1}}\mbox{cot} \left( \frac{1-\eta_{1}}{2} \pi
\right)\nonumber\\
&&\mp Y^{h_1h_2}_2  s^{-\eta_{2}}
\mbox{tan} \left( \frac{1-\eta_{2}}{2} \pi \right)
\label{RRP}\eeq
\noindent
where the upper (resp. lower) sign refers to a proton scattering with a
negatively (resp. positively) charged particle.

We now study the stability of the fit, changing $\sqrt{s_{min}}$ from
3 to 13 GeV. The number of points and the resulting $\chi^2$/d.o.f. are shown
in Fig.~1.
Clearly, the fit is bad for small energies. This is expected, as there is
no reason
then for neglecting the effect of lower trajectories.
As in \cite{six}, we need  $\, C=\pm 1\,$ meson trajectories
that  are non-degenerate, primarily because of the
constraints coming from fitting the $\, \rho \,$ parameters.
We also see
that
values of 1 or smaller for the $\chi^2$/d.o.f. can be achieved
for $\sqrt{s_{min}} = 9$ GeV. Hence,
 Regge fits are
not to be trusted below that energy.

The problem with such a high value of the minimum energy is that the
pomeron is reasonably determined, while the lower trajectories are much more
poorly fixed.
Clearly, if the fit is physically meaningful past a certain energy,
its parameters cannot depend
any longer
on its starting point.
In fact, one can see from Fig.~1 that the pomeron intercept
and couplings are stable w.r.t. $\sqrt{s_{min}}$ once we are above 8 GeV or so.
However, this is not the case for the lower trajectories: although the
$C=\pm 1$ intercepts and the $C=-1$ couplings are stable, within large errors,
the $C=+1$ couplings do depend on the minimum energy.

It is to be noted that
this problem
has to do with the definition of the error bars, as all the values of the
couplings above 9 GeV shown in Fig.~1 would lead to a $\chi^2$/d.o.f. smaller
than~1.
Furthermore, the parameters of the $C=+1$ trajectory are highly correlated
to those of the pomeron\footnote{The unrounded parameter values, the
corresponding
 dispersions, and the correlation matrices for each fit can be obtained by
request from tkachenkon@mx.ihep.su}.
A small change in the latter can produce a large
variation in the former once one is at high energy. The bottom line, however,
is that we cannot reliably determine the couplings of the $a/f$ trajectory
through the fitting procedure outlined here. This situation may change once
photon cross sections are more precisely measured at HERA and LEP.

The best we can do is
to
quote the values that we obtain for $\sqrt{s_{min}}=9$ GeV,
with the above caveats. These values are given in Table~1.
\begin{center}
{\footnotesize
\begin{tabular}{|c|c|c|c|c|c|}   \hline
$\epsilon$         &  $\eta_1 $     &  $\eta_2$ & $\chi^2$/d.o.f.&
statistics&\\ \hline
$ 0.0933\pm  0.0024$   & $0.357\pm 0.015$ & $0.560\pm 0.017$ &  1.02&383 &\\
\hline
& $pp$              & $\pi p$            & $Kp$           & $\gamma p \times
10^{-2}$& $\gamma\gamma\times 10^{-4}$\\ \hline
$X$ (mb)   & $ 18.79\pm 0.51$ & $12.08\pm 0.29$  & $10.76\pm 0.23$  &
$5.98\pm 0.17$ & $1.55\pm 0.14$\\
$Y_1 $ (mb)& $63.0\pm 2.3$      & $26.2\pm 0.74$      & $14.08\pm 0.57$      &
$11.64\pm 0.88$ & $3.9 \pm 2.0 $\\
$Y_2$ (mb) & $36.2\pm 3.2$      & $7.63\pm 0.72$       & $14.7\pm 1.3$ && \\
\hline
process&$\chi^2/N$, $\sigma_{tot}$ (N)&$\chi^2/N$, $\rho$ (N)&process
&$\chi^2/N$, $\sigma_{tot}$ (N)&$\chi^2/N$, $\rho$ (N)\\\hline
$pp$         &  1.01 (75)&    1.27  (59)&$K^+ p$      & 0.539 (22)&0.635 (7) \\
$\bar pp$    &  1.24 (35)&   0.518 (11)&$K^- p$      &  0.837 (28)&1.99  (5) \\
$\pi^+ p$    & 0.562 (24)&    2.21 (7)&$\gamma p$   &  0.624 (25)&\\
$\pi^- p$    &  1.14 (47)&  0.953 (23)&$\gamma\gamma$&   0.324 (15)&\\
\hline
\end{tabular}}
\end{center}
\begin{quote}
Table 1: The values of the parameters of the hadronic amplitude in model
$RRP$ (\ref{RRP}), corresponding to a cut off $\sqrt{s}\geq 9$ GeV,
and the values of the individual $\chi^2$ of the various processes together
with the number of points $N$.
\end{quote}

We also show the $\chi^2$ per data points and the number of data points
for each process fitted to. One can see that, as in
\cite{six}, the
$\chi^2$ is
high
for some of the sub-processes.
We have shown
in \cite{six} that this has nothing to do with the model, but rather with
the dispersion of the data. Filtering the data for these two processes did
not change the determination of the parameters. As the global $\chi^2$/d.o.f.
is
good, we do not resort here to such a procedure, as it is likely to
bias the analysis slightly.
In section 4, we shall demonstrate in another way
that this is probably
due to inconsistencies within the data, by comparing with other
parametrizations of the pomeron, and
that these high values of a few $\chi^2$ do not affect our conclusions.

The fits for
the total cross sections and $\rho$-parameters
for
$\sqrt{s}\geq 9$ GeV, extrapolated to
$\, \sqrt{s}\geq 5$~GeV,  are shown in Figs.~2 and 3.
Although the value of $\, \chi^2$/d.o.f. is bad in the low-energy region
(it goes above 2),
and thus is statistically unacceptable, the fits look deceptively satisfactory.
This shows the need for a careful statistical analysis with
physically sound criteria imposed on.

\section{The current
evidence for the simple pole ansatz}
It is not possible either to favour or to reject the simple-pole nature
of the pomeron from fits to the data. On the one hand, it
is clear that the above fit is as good as it can be once the energy is large
enough, given its $\chi^2$/d.o.f. On the other hand, as we shall see
in section 4, other fits
fare as well.
Hence, the belief that the pomeron may be a simple pole is based on the other
evidence.
\subsection{Universality}
The first requirement from
Regge theory is that the singularities
are universal,
be they
poles or cuts. Hence, the $s$ dependence of the data
has to be a combination of parts which rise or fall with energy in
a process-independent manner. Note that, in general, this does not have to
be exactly obeyed by a diagrammatic expansion of pQCD, as the hadronic
wave functions come into the calculation of the various terms in the
perturbative
expansion, and one could {\it a priori} have a small deviation from
universality \cite{CN}.
The question of whether the intercepts are
universal is also linked to the study of $F_2$ at HERA \cite{HERA}. There, for
photons with negative masses squared, it is observed that the effective
pomeron intercept, defined as
the
power of $1/x$ in $F_2(x)/x$,
seems to depend on $Q^2=-M_{\gamma^*}^2$.
It is
of interest to check whether
such a behaviour is seen on the other side of $M^2=0$.

The problem here is that very little can be said in general. One can
achieve for each process values of the $\chi^2$/d.o.f.
much smaller
than 1
if one fits to that process only.
Hence, the usual definition of errors is meaningless. For
instance, pion data have a very low
sensitivity to the pomeron
intercept. If we accepted all fits with a $\chi^2$/d.o.f. smaller than 1,
then the error bars on the various parameters
would be  much
too large to
reach any sensible conclusion. We choose here to do a partial fit,
fixing the $C=\pm 1$ meson intercepts, and letting all other parameters
free, and thus deriving errors on the pomeron intercept. As the
 $\chi^2$/d.o.f.is still small, the
errors should correspond to a change in the $\chi^2$ such as the
$\chi^2$/d.o.f.
becomes equal to 1. In order to minimize the errors, we have chosen to
include both cross sections and real parts in each fit, and kept
$\sqrt{s_{min}}=9$ GeV. We show the results of such an analysis in Fig.~4.
We have not included the $\gamma p$ data, as there is some uncertainty
regarding these. They would lead to an intercept of order 0.075 with large
error bars. Note also that in the $pp$ and $\bar p p$ cases, the
$\chi^2$/d.o.f.
is larger than 1, hence the errors correspond there to a change of 1 unit
in the $\chi^2$/d.o.f. We see that the soft pomeron intercept may be
universal, and may be independent of the target mass,
but the evidence is not overwhelming.
\begin{figure}
\centerline{
\psfig{figure=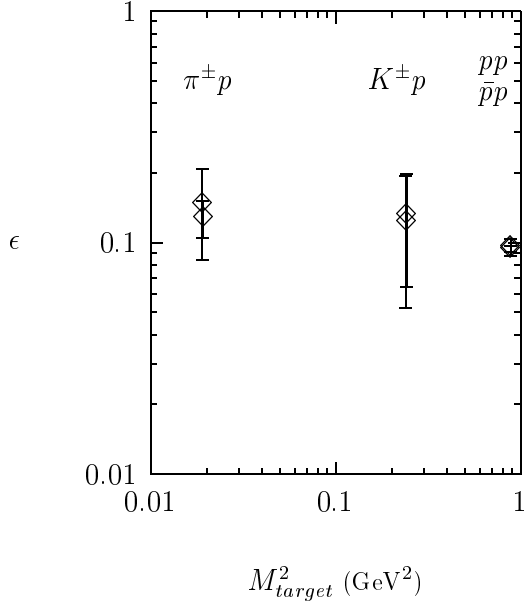,bburx=11.9cm,bbllx=4.6cm,bbury=24.9cm,bblly=16.5cm,height=8cm}
}
\begin{quote}
\caption{ The value of the pomeron intercept for three different processes.}
\end{quote}
\end{figure}
\subsection{Factorization and quark counting}
The couplings of Regge exchanges are expected to
factorize into the product of two couplings, one for each
interacting hadron.
One further and much more stringent assumption is that the pomeron
couples to single quarks as a $C=+1$ photon. The pomeron being an extended
object, this assumption is only viable for constituent quarks, and it
strongly suggests that the pomeron is a simple pole, as, otherwise, the cuts
would feel the hadronic wavefunction.
The results shown in Table 1 can be rewritten
\beq
{X_{pp}/X_{\pi p}\over 3/2}&=&1.04\pm 0.11\label{xa}\\
X_{Kp}/X_{\pi p} &=& 0.89\pm 0.05\label{xb}\\
{X_{\gamma p}\over {g_{elm}}^2\left[{1\over f_\rho^2}
+{1\over f_\omega^2}+{1\over f_\Phi^2}\right](1+\delta) X_{\pi p}}
&\approx&
{ 213.9 X_{\gamma p} \over X_{\pi p}}=1.06\pm 0.04 \label{xc}\\
{X_{pp}X_{\gamma\gamma}\over X_{\gamma p}^2}&=&0.78\pm 0.15
 \label{xd}\eeq
The first and second
relations illustrate
quark counting, the third comes from factorization
and generalized
vector-meson dominance (GVMD) \cite{GVMD},
where the contribution of off-diagonal
terms $\delta$ is expected to be about 15\%,
and the fourth is an example of factorization.
Hence, the properties of factorization and quark counting seem to hold within
10\%. However, it is clear that data from other targets would
need to be collected
at sufficiently high energy before any firm conclusion can be reached.

The quark-counting property can be summarized by rewriting the pomeron
couplings to single quarks as $1/\Lambda_u$
and $1/\Lambda_s$ for light and strange quarks, so that, for instance, we
write $X_{pp}=(3/\Lambda_u)^2$. The values of the scales thus obtained are
\beq
\Lambda_u&\approx& 0.43 {\rm \ GeV}\nonumber\\
\Lambda_s&\approx& 0.60  {\rm \ GeV}
\eeq

\noindent
It is to be noted that quark counting fails to be present for the other
trajectories. Hence, these exchanges have to probe multi-quark configuration,
whereas the pomeron seems to be coupled mainly to single quarks.

\section{The question of unitarization and alternative models}
It has been known for a long time that simple poles cannot be the
only singularities of the hadronic amplitudes, and that their existence
implies that of cuts in the complex $J$ plane.
These arise through multiple exchanges
and restore unitarity (and the Froissart-Martin bound).
These multiple-exchanges are expected to play a significant r\^ole
at the highest energies, but it is not clear
whether present data require them. The problem in studying these
is that, although one knows qualitatively what their effect will be,
nobody knows the precise form that they will take in hadronic interactions.

For instance, Donnachie and Landshoff have proposed to consider the exchange
of two pomerons as a measure of the strength of unitarization effects, others
\cite{eikonal} have used eikonal forms, or $N/D$ methods
\cite{conto}.
We
shall
not consider all the possibilities, as we shall show that
an ansatz based on an explicitly unitary answer is indistinguishable
from the simple pole fit.
\begin{figure}
\psfig{figure=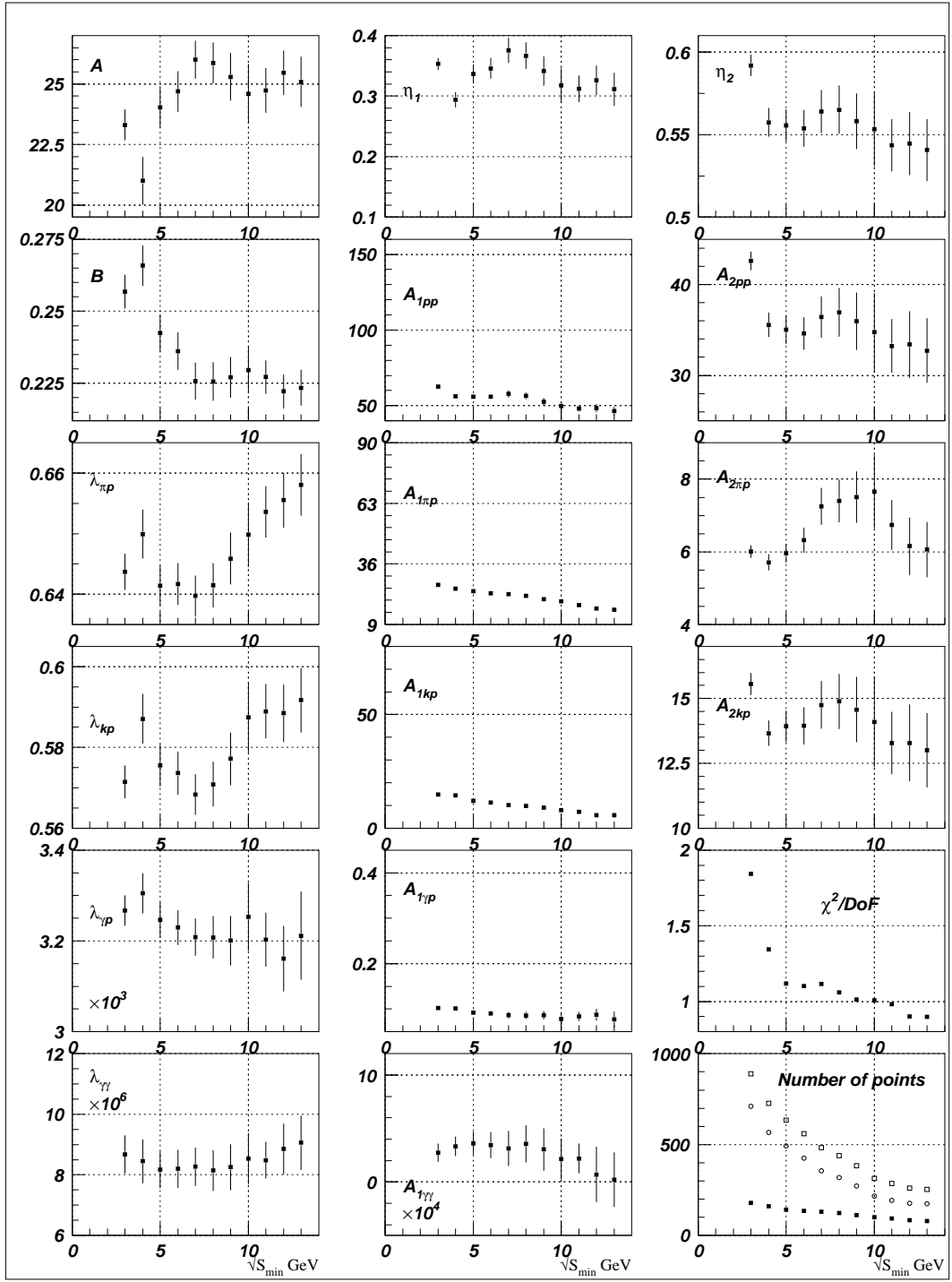,bbllx=0.7cm,bblly=1.3cm,bburx=20.3cm,bbury=28.3cm,width=14cm}
\caption{Parameters of the RRL2 model, as functions of the minimum energy
considered in the fit.}\label{fig:RRL2}
\end{figure}
\begin{figure}
\psfig{figure=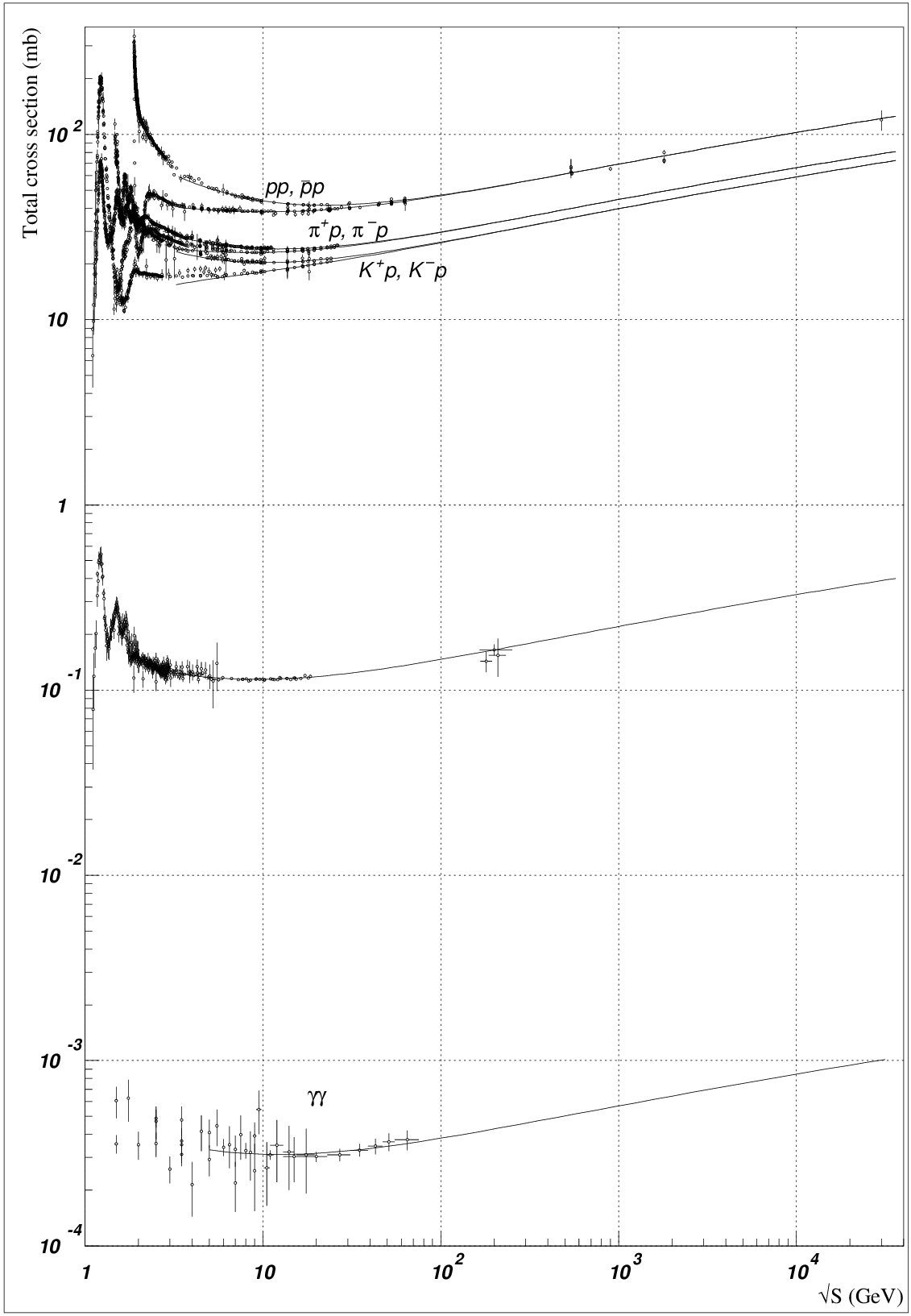,bbllx=0.7cm,bblly=1.3cm,bburx=20.3cm,bbury=28.3cm,width=14cm}
\vglue -10.5cm
\centerline{$\gamma p$}
\vglue +10.0 cm
\caption{The fit to the total cross sections from
the
parametrization RRL2.
}
\label{fig:RRL2sig}
\end{figure}
\begin{figure}
\psfig{figure=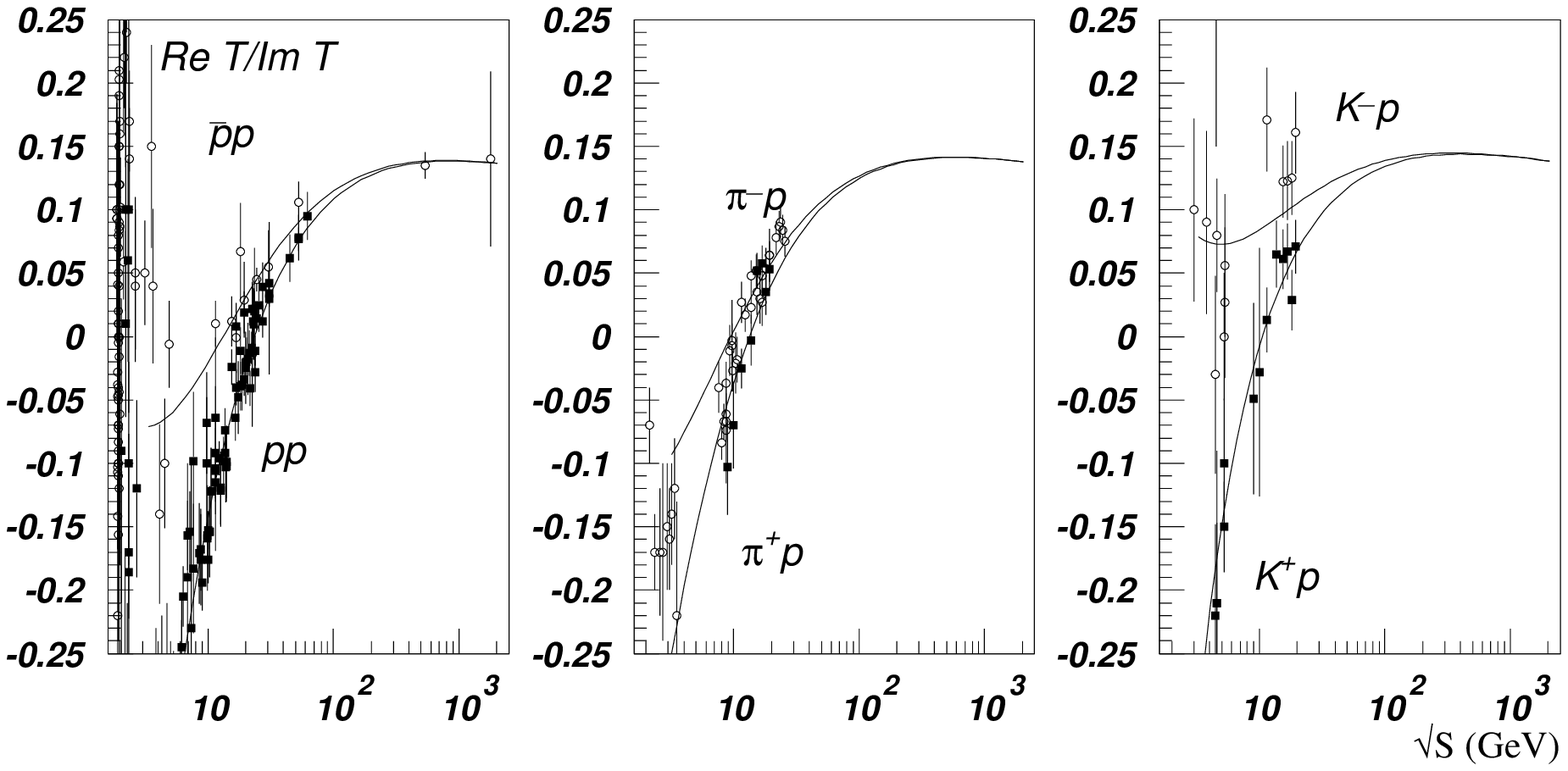,bbllx=1.3cm,bblly=10.9cm,bburx=20.cm,bbury=19.7cm,width=14cm}
\caption{The fit to the $\rho$ values from
the
parametrization RRL2.
}
\label{fig:RRL2rho}
\end{figure}

Indeed, one may assume that the simple
pole ansatz is strongly unitarized even at low $s$, and that multiple exchanges
occur
very early and turn the $s$ dependence of the cross section into
$\log^2 s$ \cite{Kang}, which would saturate unitarity, and this for energies
as low as 10 GeV. Such a form can be obtained $e.g.$ through an eikonal
formalism, although there is no justification for it in a QCD context.
Following the above logic, we have considered an amplitude of the
form
\newpage
\beq
{{\it Im A}_{h_1h_2}(s)\over s} &=& \lambda_{h_1h_2}
\left[A+B\log^2\left({s\over s_0}\right)\right]
+ Y^{h_1h_2}_1 s^{-\eta_{1}} \nonumber\\&\mp& Y^{h_1h_2}_2 s^{-\eta_{2}}\\
{{\it Re A}_{h_1h_2}(s)\over s} &=& \pi  \lambda_{h_1h_2} B
\log\left({s\over s_0}\right)
- Y^{h_1h_2}_1 s^{-\eta_{1}}\mbox{cot} \left( \frac{1-\eta_{1}}{2} \pi
\right)\nonumber\\
&&\mp Y^{h_1h_2}_2 s^{-\eta_{2}}
\mbox{tan} \left( \frac{1-\eta_{2}}{2} \pi \right)
\label{RRL2}\eeq
In order to simplify our discussion, and
to have the same number of
parameters for both fits, we set\footnote{It is possible to get slightly
better fits below $\sqrt{s}=9$ GeV$^2$ if one lets this parameter free, but
it reaches unphysical values of the order of 100 MeV$^2$ or smaller, and
the stability of the fit is not improved.} $s_0=1$ GeV$^2$.
This form, which we shall call the RRL2 amplitude,
 leads to the results shown in Fig.~\ref{fig:RRL2}.
In a manner entirely similar to the RRP case,  the fit is bad in the
region
$\sqrt{s}<9$ GeV, and again the $C=+1$ couplings are not stable. Furthermore,
it is interesting to note that this parametrization leads to fits which
are indistinguishable from the simple-pole case. Hence, all forms of
partial unitarization which lead to something between a simple pole and a
$\log^2 s$
fit cannot be distinguished on the basis of $t=0$ data alone.
We show in Figs.~\ref{fig:RRL2sig} and \ref{fig:RRL2rho} the result of such a
fit for the
cross sections and $\rho$ parameters. The parameters corresponding again to
$\sqrt{s_{min}}=9 $ GeV are given in Table 2.

\begin{center}
{\footnotesize
\begin{tabular}{|c|c|c|c|c|c|}   \hline
 $A$ (mb)        &  $B $ (mb)     &  $s_0$ & $\eta_1 $     &  $\eta_2$ &
$\chi^2$/d.o.f.\\ \hline
$ 25.29\pm 0.98$   & $0.2271 \pm  0.0071$ & $1$ (fixed)&$ 0.341\pm  0.024$   &
$0.558\pm  0.017$ & 1.01\\
\hline
 & $pp$              & $\pi p$            & $Kp$           & $\gamma p\times
10^{-2}$ & $\gamma\gamma \times 10^{-4}$\\ \hline
$\lambda$   & 1 & $0.6459\pm  0.0043$  & $0.5772\pm  0.0065$  & $0.3201\pm
 0.0055$ & $0.083\pm 0.076$\\
$Y_1 $ (mb)& $52.6\pm 2.2 $      & $20.17\pm 0.62$      & $9.00\pm 0.75 $
& $8.65\pm 0.87$ & $(3.0 \pm 2.0) $\\
$Y_2$ (mb) & $36.0\pm 3.2$      & $7.50\pm 0.71$       & $14.6\pm 1.3$ & &\\
\hline
\end{tabular}}
\end{center}
\begin{quote}
Table 2: The values of the parameters of the hadronic amplitude in model
RRL2 (\ref{RRL2}), corresponding to a cut off $\sqrt{s}\geq 9$ GeV.
\end{quote}
\newpage
Finally, it is also conceivable that unitarity is not saturated, $e.g.$
multiple exchanges of a pomeron with $\epsilon=0$ may lead to a slower
rise, in $\log s$ \cite{Kang,BKW}.
\beq
{{\it Im A}_{h_1h_2}(s)\over s} = \lambda_{h_1h_2}
\left[A+B\log\left({s\over s_0}\right)\right]
+ Y^{h_1h_2}_1 s^{-\eta_{1}} \mp Y^{h_1h_2}_2 s^{-\eta_{2}}
\label{RRL1}
\eeq
and the real part is again obtained through analyticity:
\beq
{{Re~}T_{h_1h_2}(s)\over s} =
\frac{\pi}{2} \lambda_{h_1h_2} B - {Y_{1}^{h_1h_2} s^{-\eta_{1}}}{
\mbox{cot} \left( \frac{1 - \eta_{1}}{2}\pi \right) } \mp
{Y_{2}^{h_1h_2} s^{-\eta_{2}}}{
\mbox{tan} \left( \frac{1 - \eta_{2}}{2}\pi \right) }
\eeq
The scale $s_0$ can be reabsorbed into $A$ and will be set to 1 in the
following.

\begin{figure}
\psfig{figure=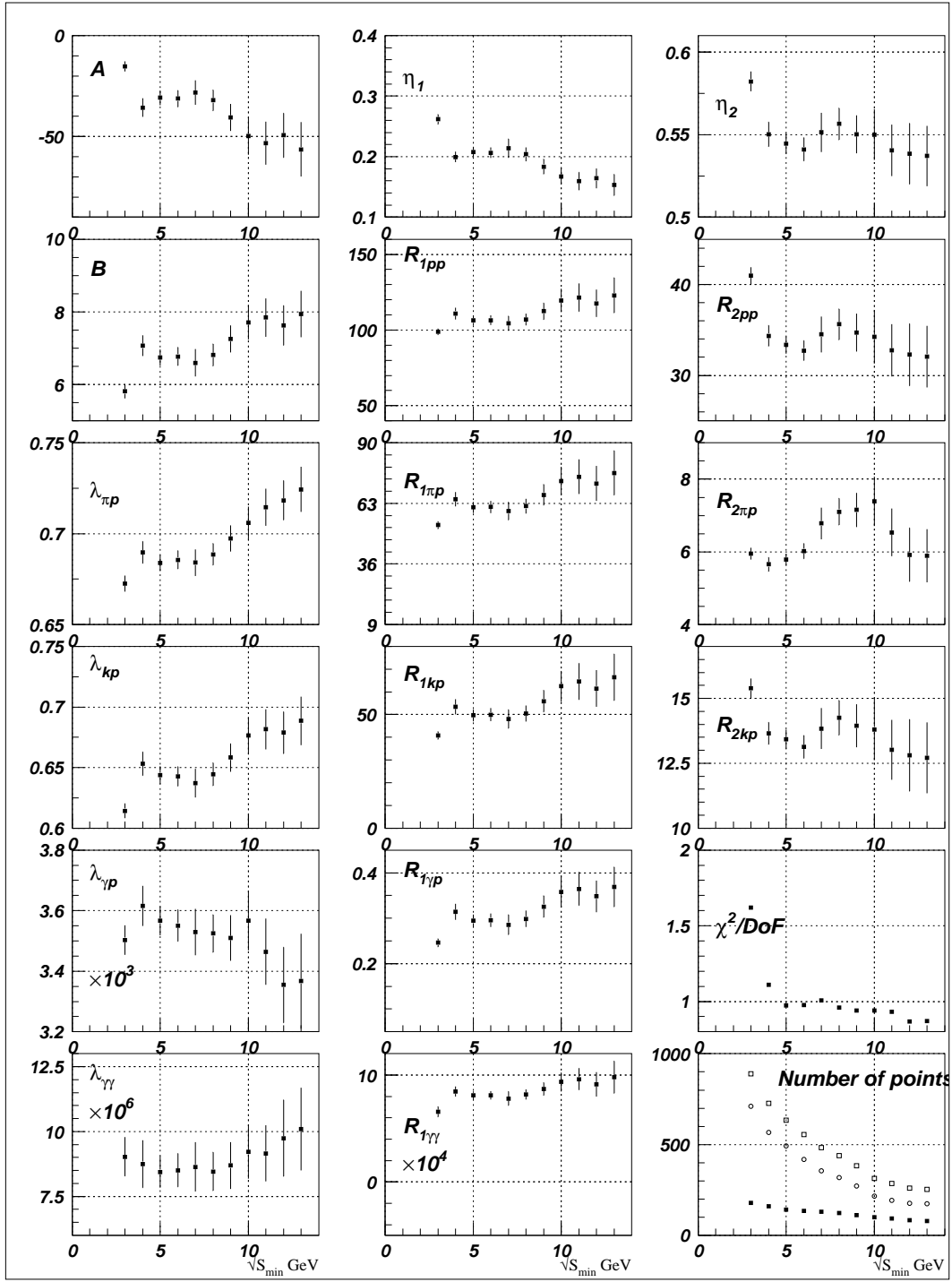,bbllx=0.7cm,bblly=1.3cm,bburx=20.3cm,bbury=28.3cm,width=14cm}
\caption{Parameters of the RRL1 model, as functions of the minimum energy
considered in the fit.}\label{fig:RRL1}
\end{figure}

\begin{figure}
\psfig{figure=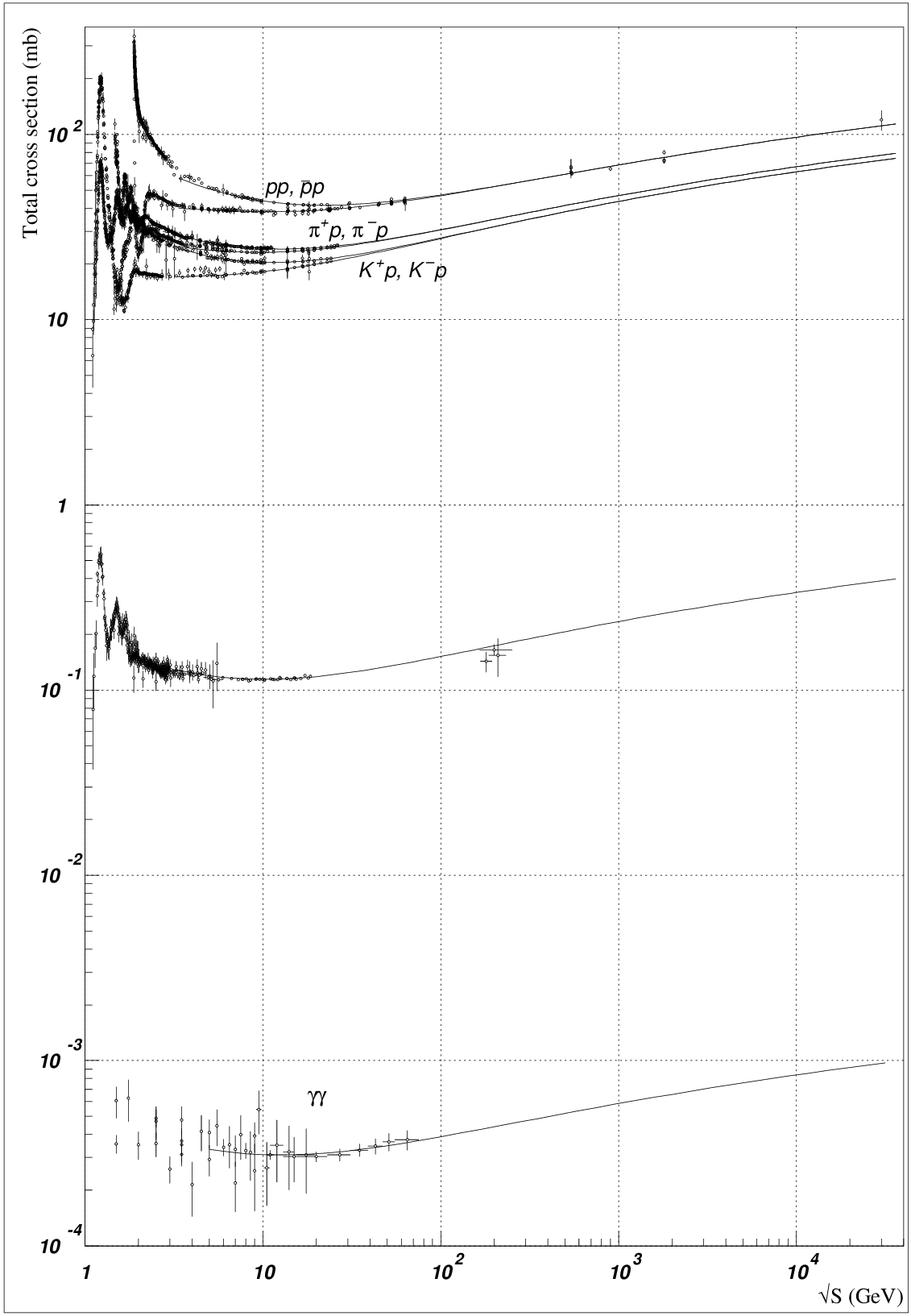,bbllx=0.7cm,bblly=1.3cm,bburx=20.3cm,bbury=28.3cm,width=14cm}
\vglue -10.5cm
\centerline{$\gamma p$}
\vglue +10.0 cm
\caption{The fit to the total cross sections from
the parametrization RRL1.
}
\label{fig:RRL1sig}
\end{figure}

\begin{figure}
\psfig{figure=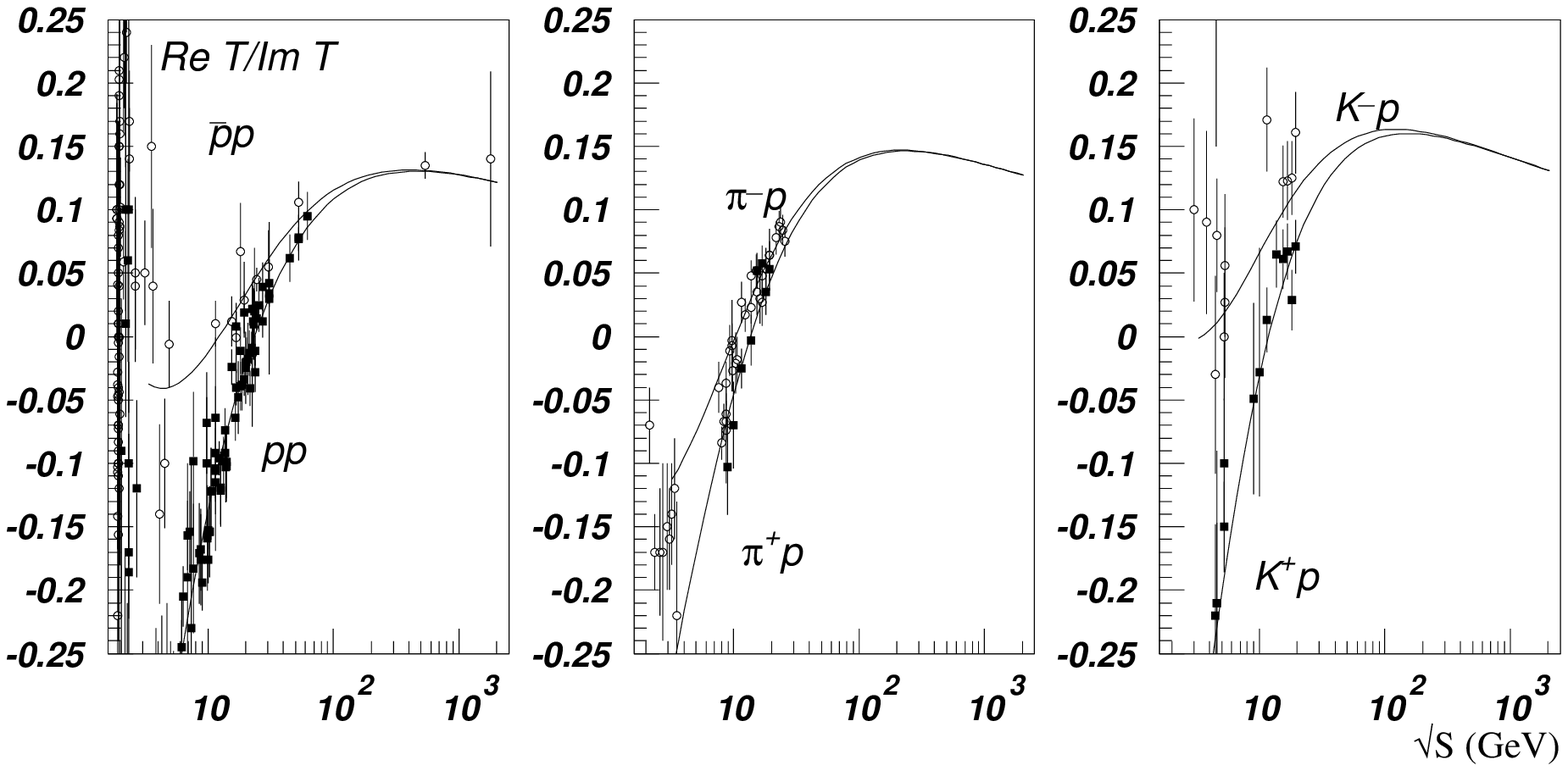,bbllx=1.3cm,bblly=10.9cm,bburx=20.cm,bbury=19.7cm,width=14cm}
\caption{The fit to the $\rho$ values from
the
parametrization RRL1.
}
\label{fig:RRL1rho}
\end{figure}

This fit leads to a slightly
better $\chi^2/$d.o.f. than the two previous ones, and to stable parameters
for $\sqrt{s}\geq 5$~GeV. We show the details of this fit, which we shall refer
to as RRL1, in Figs.~\ref{fig:RRL1}-\ref{fig:RRL1rho}, and the best values of
the parameters in Table 3. As mentioned by the E811 collaboration \cite{E811},
a logarithmic fit favours their new measurement. However, for our purpose,
we must point out that, despite the fact that the fit seems better,
the pomeron contribution becomes negative for $\sqrt{s}<12$ GeV. Hence,
one
encounters another low-energy problem, and we do not favour such fit over
the other ones for this reason.

\begin{center}
{\footnotesize
\begin{tabular}{|c|c|c|c|c|c|}   \hline
 $A$  (mb)       &  $B $ (mb)     &  $s_0$ &$\eta_1 $     &  $\eta_2$ &
$\chi^2$/d.o.f.\\ \hline
$  -30.8\pm 3.6 $   & $ 6.74\pm 0.22 $ & $1$ (fixed) & $ 0.2078\pm 0.0079  $
& $0.545\pm  0.0063$ & 0.97\\
\hline
    & $pp$              & $\pi p$            & $Kp$           & $\gamma p\times
10^{-2}$ & $\gamma\gamma\times 10^{-4}$\\ \hline
$\lambda$   & 1 & $0.6839\pm  0.0045$  & $0.6439\pm  0.0073$  & $0.3566\pm
 0.0048$ & $0.0845\pm 0.0061$\\
$Y_1 $ (mb)& $106.3\pm 2.9 $      & $61.2\pm 2.4 $      & $49.7\pm 2.5$      &
$29.4\pm 1.3$ & $8.1 \pm 3.5 $\\
$Y_2$ (mb) & $33.36\pm 0.96$      & $5.78\pm 0.16$       & $13.42\pm 0.38$ &
&\\
\hline
\end{tabular}}
\end{center}
\begin{quote}
Table 3: The values of the parameters of the hadronic amplitude in model
$RRL1$ (\ref{RRL1}), corresponding to a cut off $\sqrt{s}\geq 5$ GeV.
\end{quote}

\section{The low-energy region}
Clearly, all the fits presented here become valid only above 10 GeV or so.
Extending
of their region of validity would be an important progress, as
one would be able to compare
them with the other processes, mainly measured at lower
energies, and hence to test factorization better.

We have first tried \cite{Kangparis}  to modify the energy variable in the
fits,
and used $\tilde s={s-u\over 2}$, which is the variable predicted by
Regge theory \cite{Collins}. Note that
the use of $\tilde s$ instead of $s$ makes no difference from $\sqrt{s}=9$ GeV
onwards. It only produces significantly
better fits below $9$ GeV, but those fits are still statistically unacceptable.

We
have
also tried to implement thresholds more rigorously, and found the same
situation as
in
the $\tilde s$ case. We have also attempted to introduce lower
trajectories, but the number of parameters involved then is too high to
obtain any convincing answer. Hence, the question of the extension of the fits
to lower energies remains open.

\section{Other trajectories}
\subsection{Hard pomeron}
As we already mentioned, one of the possibilities is that the sharp rise
observed at HERA is due to the presence of another singularity, so far
undetected, which would correspond to a new kind of pomeron, called the
hard pomeron. Assuming that this is a simple pole \cite{twopoms}, one
can get beautiful fits to DIS data. However, one would expect such an
object to have some kind of manifestation in soft interactions.
Our procedure enables us to place the following 2$\sigma$ bounds on the hard
pomeron,
assuming a hard intercept of $0.4$:
\beq
{ X_{hard}^{pp}\over X_{soft}^{pp}}&<& 2\times 10^{-6}\nonumber\\
{ X_{hard}^{\pi p}\over X_{soft}^{\pi p}}&\approx& { X_{hard}^{K p}\over
X_{soft}^{K p}}< 3\times 10^{-2}\nonumber\\
{ X_{hard}^{\gamma p}\over X_{soft}^{\gamma p}}&<& 10^{-4}
\eeq
Hence, it seems that if the hard pomeron is a simple pole, it must decouple
at $Q^2\leq 0$.
\subsection{Odderon}
The exchange of a $C=-1$ trajectory \cite{odderon}
 with intercept close to 1 is needed
within the Donnachie-Landshoff model to reproduce the
large-$t$ dip in elastic
scattering.
Such an object does not seem to be present at $t=0$. Again, we can place
bounds similar to the above, but this time we allow the intercept to be as low
as 1. We then obtain
the $2\sigma$ bounds
\beq
\left|{ X_{odd}^{pp}\over X_{soft}^{pp}}\right|&<& 2\times 10^{-3}\nonumber\\
\left|{ X_{odd}^{\pi p}\over X_{soft}^{\pi p}}\right|&<& 10^{-3}    \nonumber\\
\left|{ X_{odd}^{K p}\over X_{soft}^{K p}}\right|&<&  2\times 10^{-3}
\eeq

\section{Predictions}
Finally, for each fit, we present the $1\sigma$ limits that one gets
on $\sigma_{tot}$ and
$\rho$ at current or future hadronic machines.
Clearly, one will have to wait until the results from the LHC
to
discriminate
among the various possibilities presented here. It is not even
totally
clear
whether the LHC will be able to measure total cross sections sufficiently
well within its presently approved program.

We cannot unfortunately make any firm statement on $\gamma p$ and
$\gamma\gamma$
cross sections, as both are linked. A reliable
measurement of either of these would enable
us (through factorization) to
predict the other one. At present, the published data
are not consistent or
precise
enough to reach a firm conclusion. Eq.~(\ref{xd}) shows that
factorization
would imply  higher $\gamma\gamma$ and/or lower $\gamma p$ cross sections.

\begin{center}
\begin{tabular}{|l|l|c|c|c|c|} \hline
\multicolumn{2}{|c|}{ }& {\bf RHIC}       & {\bf RHIC}
                       & {\bf Tevatron}   & {\bf LHC} \\ \cline{3-6}
\multicolumn{2}{|c|}{$\sigma^{\mbox{ab}}_{\mbox{tot}}$ [mb]}
                       & 200 [GeV]        & 500 [GeV]
                       & 2000 [GeV]       & 14000 [GeV] \\ \hline \hline
{\bf RRP} &$pp$
                       & 51.84 $\pm$ 0.18 &  60.63 $\pm$ 0.36
                       & 77.88 $\pm$ 0.87 & 111.65 $\pm$ 2.20 \\ \cline{2-6}
          &${\bar p}p$
                       & 52.03 $\pm$ 0.18 &  60.70 $\pm$ 0.36
                       & 77.89 $\pm$ 0.87 & 111.65 $\pm$ 2.20 \\ \hline
{\bf RRL2}&$pp$
                       & 52.11 $\pm$ 0.18 &  61.10 $\pm$ 0.37
                       & 78.06 $\pm$ 0.80 & 108.16 $\pm$ 1.68 \\ \cline{2-6}
          &${\bar p}p$
                       & 52.31 $\pm$ 0.19 &  61.17 $\pm$ 0.37
                       & 78.07 $\pm$ 0.80 & 108.16 $\pm$ 1.68 \\ \hline
{\bf RRL1}&$pp$
                       & 52.27 $\pm$ 0.11 &  60.97 $\pm$ 0.22
                       & 76.17 $\pm$ 0.50 & 99.90 $\pm$ 1.06 \\ \cline{2-6}
          &${\bar p}p$
                       & 52.48 $\pm$ 0.11 &  61.04 $\pm$ 0.22
                       & 76.18 $\pm$ 0.50 & 99.90 $\pm$ 1.06 \\ \hline
                       \hline
\multicolumn{2}{|c|}{ }& {\bf RHIC}       & {\bf RHIC}
                       & {\bf Tevatron}   & {\bf LHC} \\ \cline{3-6}
\multicolumn{2}{|c|}{Re/Im}
                       & 200 [GeV]        & 500 [GeV]
                       & 2000 [GeV]       & 14000 [GeV] \\ \hline \hline
{\bf RRP} &$pp$
                       & 0.125 $\pm$ 0.002 & 0.138 $\pm$ 0.003
                       & 0.145 $\pm$ 0.004 & 0.147 $\pm$ 0.004 \\
\cline{2-6}
          &${\bar p}p$
                       & 0.127 $\pm$ 0.002 & 0.139 $\pm$ 0.003
                       & 0.145 $\pm$ 0.004 & 0.147 $\pm$ 0.004 \\ \hline
{\bf RRL2} &$pp$
                       & 0.127 $\pm$ 0.002 & 0.137 $\pm$ 0.003
                       & 0.137 $\pm$ 0.002 & 0.126 $\pm$ 0.002 \\
\cline{2-6}
          &${\bar p}p$
                       & 0.130 $\pm$ 0.002 & 0.138 $\pm$ 0.003
                       & 0.137 $\pm$ 0.002 & 0.126 $\pm$ 0.002 \\ \hline
{\bf RRL1} &$pp$
                       & 0.125 $\pm$ 0.001 & 0.129 $\pm$ 0.002
                       & 0.119 $\pm$ 0.002 & 0.099 $\pm$ 0.002 \\
\cline{2-6}
          &${\bar p}p$
                       & 0.128 $\pm$ 0.001 & 0.129 $\pm$ 0.002
                       & 0.119 $\pm$ 0.002 & 0.099 $\pm$ 0.002 \\ \hline
\end{tabular}
\end{center}
\begin{quote}
{Table 4: Predictions of the $pp$ and $\bar pp$ total cross sections
and $\rho$ parameter values for future and present machines.}
\end{quote}
\section*{Conclusion}
We have shown that a simple-pole model for the soft pomeron
 produces very good fits to $t=0$ data,
once the energy is bigger than 9 GeV.
From our
updated compilation of data points, and from the 264 points above 9 GeV,
we determined the pomeron intercept to be $1.093\pm 0.003$, in agreement
with the conclusions of \cite{six}. We
have shown
that the
lower $C=\pm 1$ trajectories are non-degenerate, and have intercepts
given in Table 1. The determination of these parameters is stable and
reliable, as is that of the pomeron couplings.
We
have
also
explained
that the interplay between $C=+1$ contributions makes
the determination of the couplings of these trajectories problematic.
Further stabilization of these is needed.

Finally, we have indicated  that $t=0$ data are not sufficient to rule out
other models of forward scattering amplitudes, but the factorization
and quark counting properties, which are well respected, are difficult
to understand outside the context of simple poles.
\section*{Acknowledgements}
We thank P. Gauron for his careful check and confirmation of our fit 
   results for $\sqrt{s_{min}} = 9$ GeV and for pointing out several 
   misprints and
   mistakes in Table~1. We also thank B. Nicolescu for carefully reading 
   the e-print and for his kind proposals which made our references 
   more correct. Finally, we are indebted to J. Velasco, M.M. Block and 
especially to M. Kawasaki for pointing out that Table 4 was wrong in the
previous version of this paper.

\end{document}